\documentclass{acm_proc_article-sp}

\usepackage{enumitem}
\setlist{nolistsep}

\begin{document}
\conferenceinfo{WAHESE}{'15, May 23, 2015, Florence, Italy}
\CopyrightYear{15}
\crdata{xxx-x-xxxx-xxxx-x/xx/xx}
 
\title{Human Factors in Software Reliability Engineering}

\numberofauthors{1} 
%
\author{
%
%
\alignauthor
Maria Spichkova, Huai Liu, Mohsen Laali, and Heinz W. Schmidt\\
\affaddr{Australia-India Research Centre for Automation Software Engineering}\\
       \affaddr{RMIT University, Melbourne, Australia}\\
       \email{\{maria.spichkova, huai.liu, mohsen.laali, heinz.schmidt\}@rmit.edu.au}
}

\maketitle
\begin{abstract}
In this paper, we present our vision of the integration of human factors engineering into the software development process. 
The aim of this approach is to improve the quality of software and to deal with human errors in a systematic way.
\end{abstract}

%




\section{Introduction}

An appropriate system interface allowing correct human computer interaction is just as important as correct, 
error-free behaviour of the developed system. 
Even if the system we develop behaves in an ideal correct way (i.e., according to its requirements specification), this does not help much in the case the system interface is unclear to the user or is too complicated to be used in a proper way. As per statistics presented in \cite{Dhillon}, the human is responsible for 30\% to 60\% the total errors which directly or indirectly lead to the accidents, and in the case of aviation and traffic accidents, 80\% to 90\% of the errors were due to human. 
Thus, it is necessary to have human factors engineering as a part of the software development process.

There are many definitions of human factors, however most of them are solely oriented on human-machine operations 
in terms of system and program usability, i.e. on those parts that are seen by the (end-)user, but not by the requirements, 
specification and verification engineers. 
Nevertheless, many problems during the engineering phase are almost the same.

The fundamental goal of human factors engineering is to reduce errors, 
increase productivity and safety when the human interacts with a system, cf.~\cite{WickensHollands}. 
Engineering psychology applies psychological perspective to the problems of system design and focuses on the information-processing capacities of humans. 
It is essential to collect the error information systematically, to develop an error taxonomy and to find on this basis a solution for preventing errors of this kind. 
Thus, the fault-tolerance of software and systems engineering should be analysed not only from hardware and software side, but also taking into account human factors and Human-Computer Interaction (HCI) analysis. 

In this paper, we discuss our ideas on the software reliability improvement 
 by the integration of the  human factors engineering into the development process, 
from requirement engineering and modelling to the testing of software.  
We propose to optimise of the requirements inspection methods and the modelling methods/tools using the 
analyse of the human errors  that arise during the RE and modelling phases.
We also propose to extend the classical testing techniques by the human factor analysis, especially focusing on the human error detection and structuring. 
By predicting what types of faults are more likely to be triggered or introduced by users and/or programmers, 
we design testing strategies specifically for detecting these fault types. 
In addition, by analysis of the human usage and development history, we can prioritise the regression test cases based on the analysis results, aiming at detecting more faults earlier when the software is evolved.

\section{Background: Human Factors}

There are many applications of formal methods to analyse HCI %
and to construct user interfaces, e.g. \cite{folstad2012analysis}, as well as a number of approaches on the integrating human interface engineering with software engineering, 
e.g., \cite{Heumann2002,Constantine2003}, but the field of application of human factors to the analysis and to the optimisation of formal methods area is still almost unexplored. 
However, there are many achievements in the HCI research that could be applicable within the formal languages as well as verification and specification engineering tools. 
For example, the ideas of the usage-centered approach for presentation and interaction design 
of software and Web-based applications were introduced in \cite{ConstantineLockwood1999,ConstantineLockwood2002}.

Originally, the research of human factors and HCI was initiated and elaborated  
because of mistakes in usage and development of safety-critical systems rather than for the development of entertainment or every-day applications.
For example, one of the widely cited HCI-related accidents in safety-critical systems are the accidents involved massive radiation overdoses by the Therac-25 
(a radiation therapy machine used in curing cancer) 
that lead to deaths and serious injuries of patients which received thousand times the normal dose of radiation \cite{Miller1987,Leveson1993}. The causes of these accidents were software failures as well as problems with the system interface. 
Specifying safety-critical systems, e.g. in the automotive domain \cite{efts_book,efts}, it is not enough to use controlled languages and semiformal languages -- 
the precise formal specification is essential to ensure that the safety properties of the system really hold. 
Moreover, this should be integrated in the development methodology, preferably supported by a tool chain, cf. \cite{VerisoftXT_FMDS}. 

Speaking about human factors  
we  mostly focus on technical aspects; this idea, applied to the formal methods, is often called \emph{Engineering Error Paradigm}~\cite{RedmillRajan}. 
Human factors that are targeted by the Engineering Error Paradigm (EEP) typically include the design of human-computer interface as well as the corresponding automatisation. 
By this paradigm humans are seen as they are almost equivalent to software and hardware components in the sense of operation with data and other components, but   at the same time humans are seen as the ``most unreliable component'' of the total system. 
This implies also that designing humans out of the main system actions through automatisation 
of some system design steps is considered as a proposal for reducing risk. 
In the case of design of safety-critical systems, 
this means automatic translation from one representation kind to another one, e.g., between two formal languages or between two internal representation within some tools.

Another important view of the EEP is that human errors often occur as a result of mismatch in HCI and overestimation of physical capabilities of a person.  
With other words, we have to consider  in the design process the aspects of human performance and reliability.
In the case of models and specifications, we have to focus on clearness  and readability. 
For these reasons we have to analyse the achievements of HCI approaches to apply their ideas for %
the development of the interface between the software/verification engineers and the formal methods or tools they apply. 
The \emph{Individual Error Paradigm}~\cite{RedmillRajan} focuses on understanding the reasons why people make mistakes or commit unsafe acts, 
and then tries to eliminate those reasons. 

In the next section we discuss these aspects in the relation to requirement engineering (RE), modelling and testing phases of the software development process.

 \section{Human-Oriented  Development}
 
 The use of structured error information helps to understand the real problems in the requirements documents and  significantly improves the effectiveness of the early detection and elimination of faults in software artefacts, cf. \cite{Walia2013}.
Thus, to embed a methodology for human error analysis into the software engineering process, we have to classify the errors. 
There are a number of approaches in this field. For example,  \cite{HumanErrorTypes} introduces a
method for selection of
error types and error production mechanisms. 
A review on the strategies for the human factor taxonomies is presented in \cite{baziuk2014towards}. 

 The studies on the software requirements inspections \cite{Walia2013errorAbstr, Albayrak2014} 
have shown that the humans ability of detecting inconsistencies in requirements document 
is affected by their learning style, i.e. 
the ways with which an individual characteristically acquires, retains and retrieves the information. 
We suggest to extend these approach to the implementation and testing of the software, as the software testing teams have similar tasks and problems 
to the requirements inspection teams. 

\begin{figure}[ht!]
\begin{center}
\includegraphics[scale=0.5]{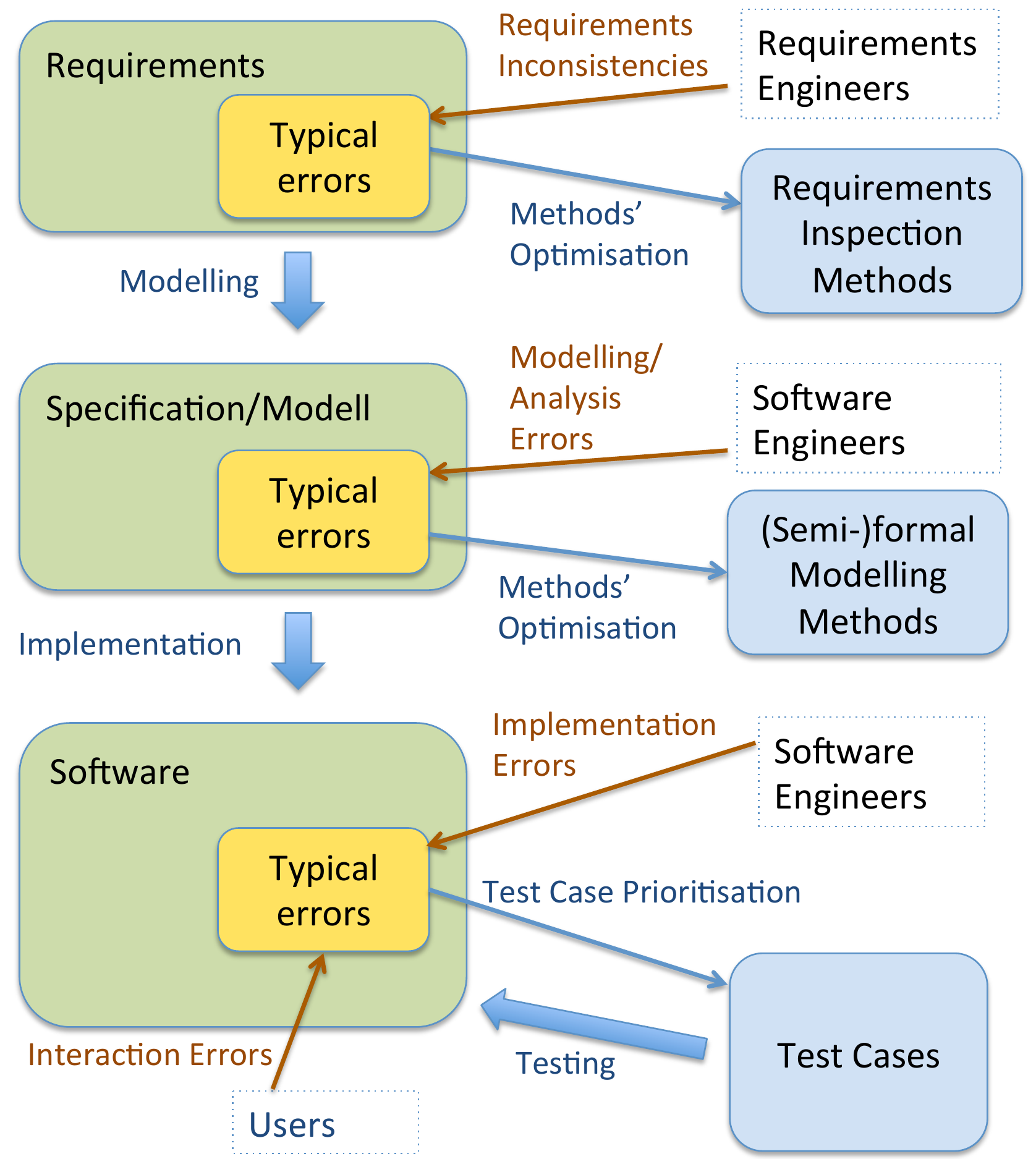}
\caption{Human-Oriented Software Development}
\label{fig:hosd}
\end{center}
\end{figure}

Figure~\ref{fig:hosd} presents the general idea of our approach on the human-oriented software development. 
A basis for optimisation of the requirements inspection methods and the modelling methods/tools  is provided by 
the analysis of typical errors that arise during the RE and modelling phases respectively.
The test case prioritisation is performed using the analysis of typical errors that arise 
(1)
during modelling and implementation phases, 
(2)
when the user interacts with the software. 
The RE, modelling and implementation phases, as well as the phase while the user interacts with the software, 
have different sets of typical errors. 
However, a taxonomy between these sets could be useful to improve the software development process. 
For example, some types of the errors which arise during the modelling phase can hardly be prevented by optimisation of the modelling methods/tools. 
Thus,  we have to specify the corresponding test cases and take this into account during the test case prioritisation.  
\\
~

\newpage
  Software reliability is usually 
defined as the probability for a certain system to perform its intended functions and operations 
for a given period of time in a specified environment. 
It is always a challenging job to achieve highly reliable software. 
There are many engineering techniques to handle software faults, the main cause of low reliability. 
The four typical families
 of software reliability engineering techniques~\cite{Lyu2007} are fault prevention, fault removal, fault tolerance, and fault forecasting. 
 \emph{Fault prevention} aims to avoid the occurrences of faults when constructing the software system (in our case, by optimisation of the methods for requirements inspections and modelling). 
 The major tasks for \emph{fault removal} are to detect and remove existing faults. Testing is the main technology for fault detection. 
 \emph{Fault tolerance} provides a mechanism that enable a system to perform normally in spite of the occurrence of faults (in our case, by analysis of the errors in HCI). 
 \emph{Fault forecasting} targets at estimating the presence of faults, which can be achieved by reliability modelling.

 \subsection{Human-Oriented RE and Modelling} 
 
 In our previous work on human factors engineering, 
 we focused on the human factor aspects in application to the formal methods used within the RE and modelling phase of a system development process \cite{DentumKeylessEntry,Spichkova2013HFFM,hffm_spichkova}.  
A software system is usually developed iteratively, the requirements specifications can be updated many timed during the development process. 
It is hard to keep up to date a readable requirements specification,  if the system model or implementation is frequently
changed during the modelling/implementation phase of the
development. This problem could be solved by appropriate
automatisation. For example, in~\cite{spichkova2013we} a (semi-)formal   
specification is generated from the system model automatically, every time the model is updated.  
 Howerver, this does not solve the problem with fault-detection in the requirements document in the first iteration of the development process, where the system model does not exists.

\subsection{Human-Oriented Testing Strategies}

\subsubsection{Human-oriented fault-based testing} 
  
 It is widely acknowledged that testing cannot ensure that a program can be fault-free, unless all possible test cases have been exhaustively executed, which is practically impossible. However, a test suite can be specifically designed such that a program can be guaranteed to be free of one particular type of fault. A testing strategy serving for such a purpose is termed as \textsl{fault-based testing}~\cite{MorellTSE1990, Tai96}. One of the oldest and also the most typical fault-based testing methods is domain testing~\cite{WhiteCohen80}. Traditionally, the input space of a program can be divided into a set of disjoint domains, each of which represents a particular path of the program. A \textsl{domain error} refers to the defect located in a decision or condition of a program. To detect domain errors, test cases are required to be selected from on and near the boundaries of every domain inside the input space. 
 
 Fault-based approach has also been widely applied into the test case selection based on Boolean specifications, cf.~\cite{FBTgbeCJ14}. 
 Boolean expressions are commonly used to describe the decisions and conditions inside a program, and different types faults can be made on the Boolean expressions, such as ``an operator is mistakenly replaced by another type of operator'', ``a term is mistakenly omitted'', etc. Various strategies have been proposed to guarantee the detection of certain fault types.

We propose that human factors be considered in designing fault-based testing aiming at enhancing the testing effectiveness. There exist infinite number of fault types; thus, it is impossible to exhaustively examine all types of faults. However, by examining the behaviours of end users, we would be able to predict which types of faults are more easily to be triggered when the software is used. In addition, we can analyse the programming styles of developers to anticipate which fault types are more likely to be introduced in the development process. Based on the analysis results, we can determine some specific types of faults that need more attention in the testing, and hence design strategies that are specifically effective in detecting these fault types. 

\subsubsection{Human-oriented test case prioritisation}

Regression testing~\cite{Rothermel1997}, which re-runs previously executed test cases, is a necessary and 
important process when changes have been made to software systems. 
Due to the rapid evolution of software, it has become critical to ensure a high effectiveness of regression testing.  
A regression testing approach means the prioritisation of the existing test cases in a certain order such that the test cases with higher effectiveness can be executed earlier. 
Various prioritisation techniques~\cite{ElbaumTSE02, Rothermel01} have been proposed to maximise the fault-detection efficiency: A prioritisation technique is considered to be more effective than the other if it can reveal more faults earlier. A typical approach to test case prioritisation is based on the information provided by the test execution history on previous versions of the software under test, such as the coverage achieved by each test case on previous versions.

In our research, we propose that the development and usage history of the software under test should also 
be investigated and applied in guiding the test case prioritisation. 
Typically, an user/programmer makes similar mistakes as usual when using/developing the software, as it is difficult to change one's working style. 
As such, if a test case $t$ reveal a fault in the original version, it is very likely that $t$ and/or those test cases similar to $t$ also cause the new version to fail. Based on the human-oriented fault history, we can further improve the effectiveness of test case prioritisation.

It is also likely that users with a similar background make similar mistakes, thus we can categorise the errors per groups of users. 
The identification of these mistakes and clarifying them not only by error types but also by the user groups, could 
(1) enable us to construct automatic models for generating test cases, 
(2) be useful for prioritising test cases based on their coverage in detecting common faults in the usage history of the software.  


\section{Conclusions}
 
In this position paper we have introduced our ongoing research on incorporation of the human factors engineering into the software development process. 
Human factors analysis can be used not only to improve quality of requirement specification and readability of software models, but also to guide various testing tasks. 
On the level of requirements specification and system modelling, we mostly focus on the readability and understandability aspects.
On the level of software testing we suggest to apply human-oriented strategies to improve the fault-detection effectiveness. Therefore, we suggest to 
analyse what types of faults are more likely to be triggered or introduced by users and/or programmers, to design testing strategies specifically for detecting these fault types. 
 
%
\bibliographystyle{abbrv}

\end{document}